Cite as:

A. Baijal, V. Agarwal and D. Hyun, "Analyzing Images for Music Recommendation," 2021 IEEE International Conference on Consumer Electronics (IEEE ICCE), pp. 1-6, USA, 2021

Note: This is the 'accepted' version of the paper. The 'published' version can be found here:

https://doi.org/10.1109/ICCE50685.2021.9427619

# Analyzing Images for Music Recommendation

Anant Baijal, Vivek Agarwal* and Danny Hyun
*Advanced R&D Group, R&D Team, Visual Display Business*
*Samsung Electronics Co., Ltd.*
Suwon-si, South Korea
{baijal.anant, danny.hyun}@samsung.com

*Abstract*— Experiencing images with suitable music can greatly enrich the overall user experience. The proposed image analysis method treats an artwork image differently from a photograph image. Automatic image classification is performed using deep-learning based models. An illustrative analysis showcasing the ability of our deep-models to inherently learn and utilize perceptually relevant features when classifying artworks is also presented. The Mean Opinion Score (MOS) obtained from subjective assessments of the respective image and recommended music pairs supports the effectiveness of our approach.

## I. Introduction

That music complements and enriches the perception of images is well researched through experiments in neuropsychology [1]. Experiencing both image and image-suited music together can result in an enriched experience for the user. For instance, a series of experiments in [2] support that happy music can make happy faces appear happier. Even though multimedia contents like dramas or movies contain manually created or curated background music, visual contents like artworks or user-generated photos are soundless. In this work, we strive to enhance the affective [3] and aesthetic experience [4] of the users by automatically recommending music playlist suiting an image's characteristics. Among various high-level characteristics, emotion is typically used as a link between the image and music modalities [5] [6] [7] [8] [9] [10] [11], given that both modalities tend to evoke emotions in humans.

For recommending image-suited music, we consider both artworks and photographs (examples can be seen in Fig. 1) and treat them differently, unlike prior work. Artworks and music appear connected not only in the common nomenclature of their movement/style (e.g. Renaissance and Impressionism in paintings and Renaissance and Impressionism in music) but also in terms of common traits that are shared between the respective movement/style [12]. Motivated from this commonality, we hypothesize that music recommended based on artwork image's movement/style (in addition to emotion) is better suited than the music recommended based on artwork's emotion alone. In other words, we hypothesize that the characteristics of music suiting an artwork image of a certain movement/style may be different from that of the music suiting an artwork of a different movement/style, even though these images may evoke a common emotion. In case of photographs, we utilize image's emotion alone (similar to [6] [10]) for recommending music. We

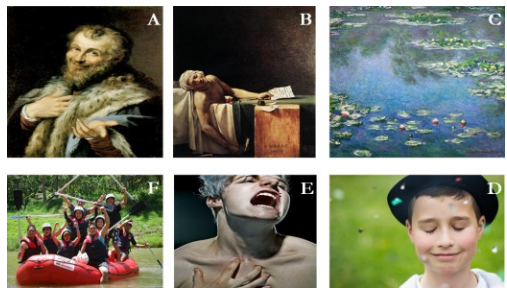

(Clockwise, starting from top-left)
A: [+V, +A, Baroque] RV 113 / Concerto for strings & b.c., Vivaldi *(Happy-Baroque music)*
B: [-V, -A, Classical] The Requiem in D minor, K. 626, Mozart *(Sad-Classical music)*
C: [+V, -A, Impressionist] A la maniere de Borodine M. 63/1, Maurice Ravel *(Calm-Impressionist music)*
D: [+V, -A] Fire and Rain, James Taylor *(Calm Music)*
E: [-V, +A] Bad Listener, Beartooth *(Intense Music)*
F: [+V, +A] Connection, OneRepublic *(Happy Music)*

**Fig. 1:** Exemplar image-suited recommended music retrieved via an online-streaming music service [20] based on results of automatic image classification (text in italics denotes keywords used by our music metadata engine for querying the music service provider). Our deep-learning based image classification models can automatically distinguish between photographs and artworks, classify artworks into their respective movement/style (e.g. impressionist, classical, baroque etc.) and determine the valence [V; positive (+V) or negative (-V)] and arousal [(A); high (+A) or low (-A)] to characterize image's emotion. Photograph images from [23]. Artworks images (in public domain) from [21]. (Images best seen in color)

categorize emotions according to the circumplex model [13], a dimensional model that groups emotions using a two-dimensional valence-arousal (V-A) affective plane wherein valence characterizes whether the emotions are positive (+V) or negative (-V) and arousal characterizes the intensity of the respective emotion as high (+A) or low (-A). Instead of using hand-crafted features [16], we utilize deep neural networks for classifying images. Deep-learning techniques have shown promise in relevant image classification tasks such as image emotion classification [14] [15] [17] and image style classification [18] [19] [33]. Based on results of image analysis, music is retrieved via our music metadata engine that queries an online music-streaming provider [20] with relevant keywords. While there are several ways to access music on latest devices, by using an online music-streaming service we can provide the users a wider variety of music that is continuously updated. This is in contrast to other existing studies wherein music is retrieved from a pre-stored database based on matching common characteristics between images and music.

---

* Affiliation at the time of contributing to this paper

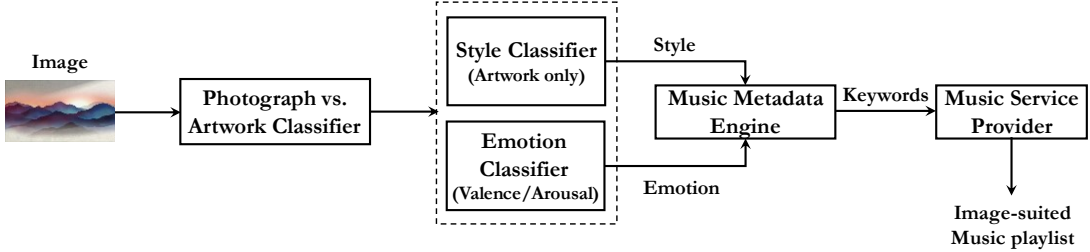

**Fig. 2:** Overview of our approach

Our key original contributions are *threefold*: *I)* In case of artworks, we propose to utilize the artwork's movement/style in addition to its emotion for recommending music (in case of photographs we utilize image's emotion alone and verify the importance of emotion as a mediator between image and music modalities [22]). *II)* Our deep-learning models for image analysis outperform the objective performance of state-of-the-art image-emotion classification (both artworks and photographs) and artwork-style classification tasks thereby setting new benchmarks for the research community. *III)* We present an illustrative analysis showing the ability of our deep-models to inherently learn the perceptually relevant features useful in classifying artworks. The Mean Opinion Score (MOS) obtained from subjective assessment of image-music pairs supports the effectiveness of our approach to analyzing images for music recommendation.

## II. OVERVIEW OF OUR APPROACH

The overview of our approach (patent-pending [34]) is illustrated in Fig. 2. We implement our work on a multimedia device (specifically, a smart television) capable of displaying images and playing music via the online-streaming service.

### A. Image classification

Our deep-models perform four required automatic image classification tasks: i) distinguishing between artworks and photographs; ii) classifying photographs based on emotion; iii) classifying artworks based on emotion; iv) classifying artworks based on style. The first step in our algorithm is automatic classification of the input image into artwork or photograph. In the second step, the respective deep-learning model (separate emotion models are used for artwork and photograph) classifies the image into an emotion category (characterized by valence and arousal). The artwork is further classified into its movements/styles (e.g. classical, impressionism, abstract etc.).

### B. Music metadata engine

Our music metadata engine contains a set of pre-defined music metadata/keywords— emotion keywords for photographs and emotion-style keywords for artworks.

*a) Artworks:* The music metadata/keywords for artworks are designed such that they encompass both image's emotion and style. For example, when an artwork is classified as classical (style) and happy (emotion), the keyword characterizing music is 'happy classical'. In other words, the metadata/keywords are intended to retrieve music having similar/same style-emotion as that of the artwork image. Some exemplar style keywords include 'baroque', 'classical', 'renaissance', 'impressionist', 'abstract', 'romanticism' etc.

*b) Photographs:* We recommend music evoking the similar/same emotion as that of the photograph (similar to [6] [10]). Accordingly, based on the circumplex model [13], any emotion keyword can be selected from the emotions associated with the respective quadrant into which the photograph is classified. For example, if the photograph is classified into [– V –A] emotion quadrant, then the emotion metadata/keywords sent to the music streaming service can be at least one of 'calm', 'relaxed', 'sleepy', 'contentment' etc. to retrieve music characterizing image's emotion.

## III. DATASETS

### A. Photograph datasets

*a) Deep Emotion [23]:* This dataset contains about 23,000 images divided across eight emotion categories – Amusement, Awe, Anger, Contentment, Excitement, Disgust, Fear, and Sadness. We re-group these emotions as shown in Table I. It is noted that we exclude the emotions amusement and awe from the arousal category, as the images from both categories appear to belong to both high and low arousal classes (as also inferred from [24]).

TABLE I. RE-GROUPED VALENCE-AROUSAL EMOTION CLASSES

| V/A | Class | Emotions |
|---|---|---|
| Valence | Positive | Amusement, Contentment, Awe, Excitement |
| Valence | Negative | Anger, Disgust, Fear, Sadness |
| Arousal | High | Anger, Excitement, Disgust, Fear |
| Arousal | Low | Contentment, Sadness |

*(b) WEBEmo [17]:* This is the largest image-emotion dataset available comprising of ~268K weakly labelled images divided across 25 emotion categories, collected though web-scraping. We utilize only binary valence labels for our purpose.

*(c) UnbiasedEmo [17]:* This dataset is an unbiased emotion test set consisting of 3045 images. Results on this dataset can support the generalizability of the trained emotion model as opposed to other datasets which are found to be biased to a certain degree [17].

### B. Artwork datasets

*a) WikiArtSubset:* For artworks, we use the publically available Wikiart dataset [21] containing about 80,000 images.

Although it contains style labels for artworks, emotion labels are not directly available. To train our models for emotion classification on artworks, we labelled a randomly selected subset (approximately 6000 images) of these artworks internally. We call it WikiArtSubset.

*(b) WikiArt Emotions Dataset [25]*: This is a dataset of 4105 pieces of art (mostly paintings) labelled across 20 emotion categories, six of which are mixed/neutral emotions. Moreover, each image is mapped to one or more emotions, making it difficult to train a hard boundary emotion classifier on this dataset. For our purpose, we filter out images that belong to purely positive valence (belonging to one or more positive emotions only) or purely negative valence category. As a result, we obtain 1940 images (1484 positive, 456 negative).

## IV. DEEP MODELS

For classification tasks, we train two kinds of deep models – one model without pre-training and one with pre-trained weights. For classification tasks mentioned in Table II, we evaluate both kinds of models whereas for the remaining tasks tasks we evaluate the models with pre-training only (results in Table III). The details of these models are mentioned below.

### A. Models without pre-training (as baseline)

We create this model to get baseline accuracy on our classification tasks where previous results for comparison cannot be found in the literature. The network comprises of 4 convolutional layers followed by one fully-connected hidden layer (512 units) and one output layer consisting of 2 softmax units for binary classification. The number of filters in each convolutional layer is 32, 64, 128 and 256 respectively, each of which is a 3x3 in kernel size. Also, a stride of 2x2 is used in each convolutional layer except the first one for which stride is 1x1. We use batch-normalization after each layer in the model and 'relu' activation function in all the hidden layers.

### B. Models with pre-training

We use ResNet50 architecture [26] and initialize it with pre-trained weights on ImageNet dataset. The last layer of standard ResNet50 (containing 1000 units) is removed and replaced by a softmax layer with two units for binary emotion classification (valence and arousal) and 'n' units in case of style classification, where n is the number of styles.

## V. TRAINING METHODOLOGY

### A. Learning rate policy

All the models are trained using backpropagation with 'Adam' optimizer [27] to update the neural network weights. In conjunction, we also made use of cyclic learning rates (CLR) [28] and stochastic gradient descent with warm restarts (SGDR) [29] as our learning rate policy for all our models (including the baseline models). Specifically, for each dataset, we calculate optimal range of learning rates using the LR range test defined in [28]. During training, the learning rate is then varied between these bounds cyclically following SGDR strategy [29]. We also employ learning rate decay between cycles (typically between 0.8 and 0.9) along with increasing the cycle length in each successive cycle.

### B. Fine-tuning ResNet50

To train ResNet50 model, first, we freeze all the layers of the model and fine-tune only the last layer for some epochs (2 - 3 epochs). Next, we start unfreezing the layers from the end of the network one by one. We unfreeze one additional convolutional layer at a time and fine-tune it for a few epochs (1~2). We repeat this procedure for 5~10 convolutional layers and use early stopping for model convergence. It is noted that we use data augmentation (horizontal image flip, zoom, rotation) throughout our training to avoid over-fitting. We use gradient descent with warm restarts (SGDR) [30] as our learning rate policy for all our models (including the baseline models).

## VI. OBJECTIVE RESULTS AND ANALYSIS

We use objective tests to analyze the performance of image-analysis algorithms. All the obtained results are listed in Table II and Table III (subjective assessment for image-suited music recommendation can be seen in Sec. VII).

### A. Artwork vs. Photograph classification

We achieve a 95.7% accuracy with fine-tuned ResNet50 model.

### B. Image Emotion classification

For each type of image (photographs and artwork), we create separate emotion models. Each emotion model has two sub-models: Valence and Arousal. Both valence and arousal tasks are binary classification tasks.

*a) Photographs: Testing on Deep Emotion [23]:* We outperform existing state-of-the-art results [17] on binary valence classification achieving 85.9% on this dataset. We also report first results on binary arousal classification (80.1%). For comparison purposes with the benchmark [23], we train a model on photograph dataset containing eight categories of emotions - Anger, Awe, Contentment, Disgust, Excitement, Fear and Sadness. We surpass the results of the benchmark paper as shown in Table II.

*b) Artworks: Testing on Wikiart Emotions [25]:* For artwork emotion classification, there is no prior benchmark results to the extent of our knowledge. We report first results on Wikiart emotions dataset. We achieve a 75.77% accuracy on valence classification with our model trained on WikiArtSubset. Due to the nature of labels in this dataset (such as 'calmness and 'ecstasy' being clubbed together in the same category), arousal classification cannot be performed.

*c) Photographs: Testing on UnbiasedEmo [17]:* We also evaluate our models on UnbiasedEmo dataset, which is claimed to be the most unbiased publicly available emotion dataset [17]. We test on this dataset in two ways: one in which a separate model is trained on UnbiasedEmo with some pre-trained model as feature extractor and the other where no model training is performed and results are obtained directly from pre-trained model. The model used in the former (on top of feature extractor) comprises of two dense layers of 64 and 32 neurons with dropout 0.3. We use 80% images for training and 20% for testing. For this experiment, we use two pre-trained models - one trained on Deep Emotion dataset, and the other trained on WEBEmo data. From Table III, our best result (WEBEmo as pre-trained model and fine-tuned on UnbiasedEmo) manages to surpass the benchmark results [17] by approximately 9 percentage points (74.27%— benchmark versus 83.45% — our results). Our results also support the conclusion by [17] that a model trained on WEBEmo is more generalizable than the one on Deep Emotion dataset.

TABLE II. ACCURACY OF DEEP-MODELS ON DIFFERENT IMAGE CLASSIFICATION TASKS

| TEST Dataset | TRAINING Dataset | Classification Task | ^: State-of-the-art *: Baseline | OUR RESULTS (Fine-tuned-ResNet50) |
|---|---|---|---|---|
| Wikiart Emotions [25] (filtered version) | WikiArtSubset | Valence (artworks) | *53.76% | **75.77%** *(first-ever reported results on [25])* |
| WikiArtSubset | WikiArtSubset | Valence (artworks) | *79.4% | **89.0%** |
| | | Arousal (artworks) | *78.5% | **88.6%** |
| Deep emotion [23] | Deep Emotion (refer Table I) | Arousal (photographs) | *60.8% | **80.1%** *(first-ever reported results)* |
| | | Valence (photographs) | ^ 84.81% By [17] | **85.9%** *(new state-of-the-art)* |
| | Deep emotion (8 classes) | 8 Emotions (photographs) | ^ 58.3%, By [15] ^ 61.13%, By [17] | **61.3%** *(new state-of-the-art)* |
| UnbiasedEmo [17] | WebEmo + UnbiasedEmo (fine-tuning) | Valence (photographs) | ^ 74.27% By [17] | **83.45%** (refer Table III) *(new state-of-the-art)* |
| Wikiart [21] | Wikiart (27 style classes) | Style (artworks) | ^ 54.50% By [33] | **58.42%** *(new state-of-the-art)* |

TABLE III. ACCURACY ON UNBIASEDEMO DATASET [17]

| | Model Pre-training On | Finetuned on UnbiasedEmo | Accuracy |
|---|---|---|---|
| *Benchmark* [31] | WEBEmo | Yes | 74.27% |
| **Our Results** | WEBEmo | No | 76.55% |
| | WEBEmo | Yes | **83.45%** |
| | Deep Emotion | No | 67.40% |
| | Deep Emotion | Yes | **77.20%** |

## C. Artwork style classification

For artworks, we train a classification model to predict its movement/style. Original dataset of 80,000 Wikiart paintings comprises of 27 style categories (e.g. High Renaissance, Early Renaissance, Impressionism, Expressionism, Cubism, etc.). We train our model on 27 categories and surpass the state-of-the-art performance by achieving 58.42% accuracy (Table II).

## D. Illustrative analysis

We find that for an image, perceptually important features such as its color profile, geometry of the shapes present in it, as well as its semantics are inherently captured and utilized by the deep-models for image emotion classification.

*a) Importance of color:* Color is seen as a primary factor by many psychological studies to evoke a particular kind of emotion [30] [31] [32]. For example, bright colors such as red, orange are known to typically induce intense emotions as compared to blue or turquoise which elicit calmer emotions. We observe similar behavior in working of our deep-model. For instance, although the semantic meaning of all images in Fig. 3 is similar (all paintings represent mountain scenes), the color of the background appears to play a decisive role in deep-model's classification of the images' into high and low arousal levels, similar to perceptual phenomenon. Similar perceptually-aligned behavior of the trained deep-models is observed along the valence plane wherein images having vivid and bright colors are classified onto the positive side of the valence axis whereas presence of grays and blacks results in the image being classified onto the negative side of the valence axis (sadness, gloom, depression etc.).

*b) Importance of Geometry:* We illustrate in Fig. 4 how the geometry of the shapes present in an image plays a key role in image classification. For abstract artworks, we can observe that images with predominantly ragged and unsymmetrical shapes such as crowded and randomly directed lines tend to be classified onto the high arousal plane (e.g.: anger, fear) whereas images with predominantly symmetrical and softer shapes such as circles and evenly spaced squares tend to be classified onto the low arousal plane of emotion.

*c) Importance of Semantics.* Though highly intuitive for humans, capturing semantics of an image is not generally

considered an easy task for machines. Yet, our deep-model can automatically identify the semantics necessary for classifying emotion. As an example shown in Fig. 5, the model identifies the respective emotion quadrant by automatically taking into account the facial expression depicted in the image, especially when color profiles or geometry prove to be indecisive.

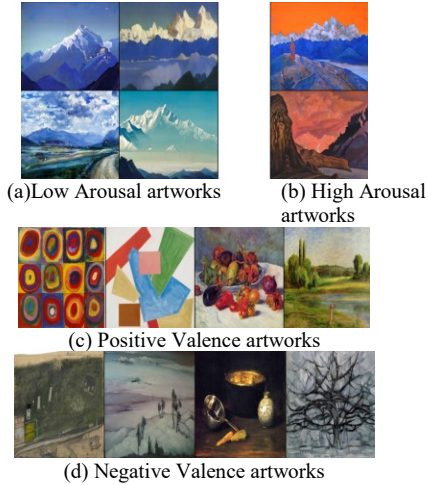

(a) Low Arousal artworks    (b) High Arousal artworks

(c) Positive Valence artworks

(d) Negative Valence artworks

**Fig. 3:** Importance of Color in predicting emotion. Artworks (in public domain) from [21] (Best viewed in color)

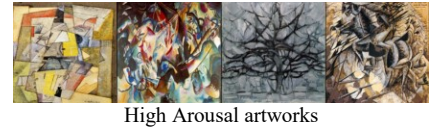

High Arousal artworks

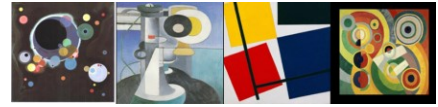

Low Arousal artworks

**Fig. 4:** Importance of Geometry in predicting emotion. Artworks (in public domain) from [21] (Best viewed in color)

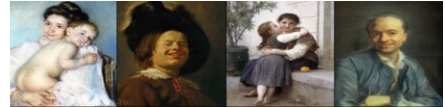

Positive Valence artworks

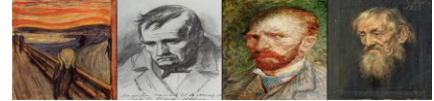

Negative Valence artworks

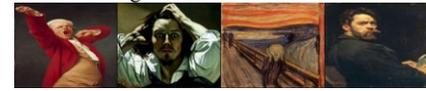

High Arousal artworks

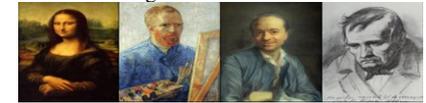

Low Arousal artworks

**Fig. 5:** Importance of semantics in predicting emotion. Artworks (in public domain) from [21] (Best viewed in color)

## VII. SUBJECTIVE ASSESSMENT AND ANALYSIS

We design a subjective test to compare the following three cases.

• *Our approach:* Music recommended by considering both emotion and style of artwork image— to this end, both style and emotion keywords are used together (see Sec. II) for querying and retrieving respective music via Spotify [20].

• *Existing approach:* Music recommended by considering only emotion of the image— to this end, only an emotion keyword belonging to the same quadrant of the circumplex model as the image is used for querying and retrieving respective music via Spotify. However for photographs, we select two test music pieces as style is not relevant in this case.

• *Baseline approach:* Music recommended such that its emotion 'mismatches' with that of the emotion of the image— to this end, we retrieve music by querying with an emotion keyword belonging to a different quadrant of the circumplex model than that of the image.

Our test design is as follows: Each human subject is presented with 16 test images (eight photographs and eight artworks), with equal number of images (four images) for each valence-arousal quadrant (+V +A; +V –A; -V –A; +V -A). The images for each quadrant were chosen at random. For each image, three aforementioned music pieces (each of length 15 seconds) are presented to the human subject. The sixteen images are presented in random order to the subjects and for each image the three test music pieces are presented in random order— the nature of the test is blind. The overall subjective test typically lasts between 15 and 18 minutes depending on the subject. It is noted that test music pieces are retrieved from a single Spotify account irrespective of subjects' personal taste and preference. A test image is shown to the subject and the subject is asked the following for each of the three music pieces for the respective image: "Please rate the suitability of the music for this image." For the subjective assessment, we use a Mean Opinion Score (MOS) rating on a scale of 1 to 5, where the rating scale is defined as: 1- Bad, 2-Poor, 3- Fair, 4- Good and 5-Excellent. The subjects are free to listen to the respective music with the respective image any number of times.

Total 74 subjects (58 males and 16 females) of various ages (between 20s and 60s) participated in our subjective test. The same subjects evaluated all image-music test pairs. In total, 3552 test responses were recorded. For artworks, when recommending music using both emotion and style (proposed approach), we obtain an MOS of 3.75 (see Table IV). However, when recommending music using artwork's emotion alone, the MOS falls to 3.45. We conduct a paired t-test (results in Table V) in order to statistically verify the MOS ratings. We take the mean of scores for all the eight test artwork images for each of the three test (music) cases for each participant. The null hypothesis (H0) assumes that the true mean difference between emotion-style (C) and emotion (B) based subjective ratings is zero while the alternative hypothesis (H1) assumes that the true mean difference is not zero. It can be inferred from Table V that the aforementioned null hypothesis is rejected in favor of alternative hypothesis thereby confirming the efficacy of our approach that proposes to utilize both style and emotion of the artwork when recommending suitable music. Moreover, the respective MOS values (see columns (A) and (B) in Table IV) confirm the importance of emotions as a mediator in cross-matching image and music modalities for both artworks and photographs, and this is further statistically verified by the respective *p*-values (see row 'A vs. B' in Table V). In other words, the participants prefer experiencing recommended music that evokes emotion matching that of the image, for both artworks and photographs.

TABLE. IV SUBJECTIVE PERFORMANCE AS JUDGED BASED ON MEAN OPINION SCORE (MOS) RATINGS

| Number of Participants = 74 | (A) Baseline Approach *Mismatched Emotion* | (B) Existing Approach *Matched Emotion* | (C) Our Approach *Matched Emotion & Style* |
|---|---|---|---|
| Artworks | 2.66 | 3.45 | **3.75** |
| Photographs | 2.48 | **3.72** | (style not relevant) |

TABLE. V RESULTS OF PAIRED T-TESTS (Confidence level: 95%) (Also see Table IV)

| Paired t-tests Test Cases | Photographs | Artworks |
|---|---|---|
| **B vs. C** (see Table IV) | (style not relevant) | $H = 1$ $t(73) = 6.65$, $p < .001$ |
| **A vs. B** (see Table IV) | $H = 1$ $t(73) = 20.42$ $p < .001$ | $H = 1$ $t(73) = 16.44$, $p < .001$ |

## VIII. CONCLUSIONS

We present a novel method to address the problem of analyzing images for suitable music recommendation by treating artworks and photographs differently. We hypothesize that recommending music aligned with the respective movement/style of artworks (in addition to artwork's emotion) is more suitable than recommending music based on emotion alone and verify this hypothesis using subjective assessment coupled with statistical analysis. We utilize deep-learning based models and achieve state-of-the-art results on automatic image emotion and image style classification tasks. Our illustrative analysis shows that these deep-models inherently learn, and make use of, perceptually important features (color, geometry and semantics) to perform artwork emotion classification.